\documentclass[11pt]{article}
\usepackage[margin=1in]{geometry}
\usepackage{graphicx}
\usepackage{lineno}

\newcommand{\nuebar}{\ensuremath{\overline{\nu}_{e}}}
\newcommand{\uFive}{$^{235}$U}
\newcommand{\uEight}{$^{238}$U}
\newcommand{\pNine}{$^{239}$Pu}
\newcommand{\pOne}{$^{241}$Pu}

\def\beq{\begin{equation}}
\def\eeq#1{\label{#1}\end{equation}}
\def\eeqn{\end{equation}}

\def\beqa{\begin{eqnarray}}
\def\eeqa#1{\label{#1}\end{eqnarray}}
\def\eeqan{\end{eqnarray}}

\let\bar=\overbar

\def\Dslash{\not{\hbox{\kern-4pt $D$}}}
\def\dslash{\not{\hbox{\kern-2pt $\del$}}}

\def\msb{{\bar{\ssstyle M \kern -1pt S}}}

\def\Title#1{\begin{center} {\Large {\bf #1} } \end{center}}
\def\Author#1{\begin{center} {\normalsize {\sc #1} } \end{center}}
\def\Institution#1{\begin{center} {\normalsize {\it #1} } \end{center}}
\def\Abstract#1{\noindent {\normalsize {\bf Abstract:} {\normalfont #1}}}
\def\Conference{\vspace{4mm}\begin{raggedright} {\normalsize {\it Talk presented at the 2019 Meeting of the Division of Particles and Fields of the American Physical Society (DPF2019), July 29--August 2, 2019, Northeastern University, Boston, C1907293.} } \end{raggedright}\vspace{4mm}}

\begin{document}

\Title{Measurement of the Reactor Antineutrino Spectrum from $^{235}$U Fission using PROSPECT}

\Author{Pranava Teja Surukuchi \\ \textit{(for the PROSPECT Collaboration)}}

\Institution{Wright Laboratory, Yale University, New Haven CT - 06511, USA}

\Abstract{PROSPECT is a short-baseline reactor antineutrino experiment designed to search for short-baseline sterile neutrino oscillations and perform a precise measurement of $^{235}$U reactor antineutrino spectrum from the High Flux Isotope Reactor at Oak Ridge National Laboratory. This measurement probes our understanding of recent anomalous results observed in reactor antineutrinos. PROSPECT uses a $\sim$4-ton optically segmented, $^{6}$Li-loaded liquid scintillator detector with high light yield, world-leading energy resolution, and excellent pulse shape discrimination. The latest antineutrino spectrum measurement results from $^{235}$U fissions at HFIR are reported.}

\Conference 

\section{Introduction}

In the last decade, there has been a resurgence in short baseline experiments motivated by the discrepancies between predictions and experimental measurements~\cite{Mueller11,Huber11}.
The average flux measured by the reactor experiments are lower by $\sim$6\% compared to the predictions.
This \textit{flux anomaly}~\cite{Mention:2011rk} could be due to incorrect flux predictions or due to the existence of an additional sterile neutrino state at $\sim$1eV$^2$.
Precision reactor neutrino experiments Daya Bay, Double CHOOZ, and RENO \cite{An:2016srz,Abrah_o_2017,choi2016observation} built to measure $\theta_{13}$ have also noticed deviations from the predictions in the shape of the spectrum primarily in 5-7 MeV neutrino energy region. 
This anomaly cannot be explained by a sterile neutrino but rather points to inaccuracies in spectrum predictions arising from mistakes in databases or errors made in \textit{conversion} process.
All the experiments that have measured the spectral deviations have sampled the \nuebar~from a Low Enriched Uranium~(LEU) reactors which generates the \nuebar s from four primary isotopes namely \uFive, \uEight, \pNine, and \pOne.
By measuring the \nuebar~spectrum from a Highly Enriched Uranium~(HEU) nuclear reactor where the \nuebar~are primarily sampled from a single isotope the isotopic source of discrepancy can be experimentally disentangled.

\section{PROSPECT Experiment}
\begin{figure}[htb]
\centering
\includegraphics[width=0.6\textwidth]{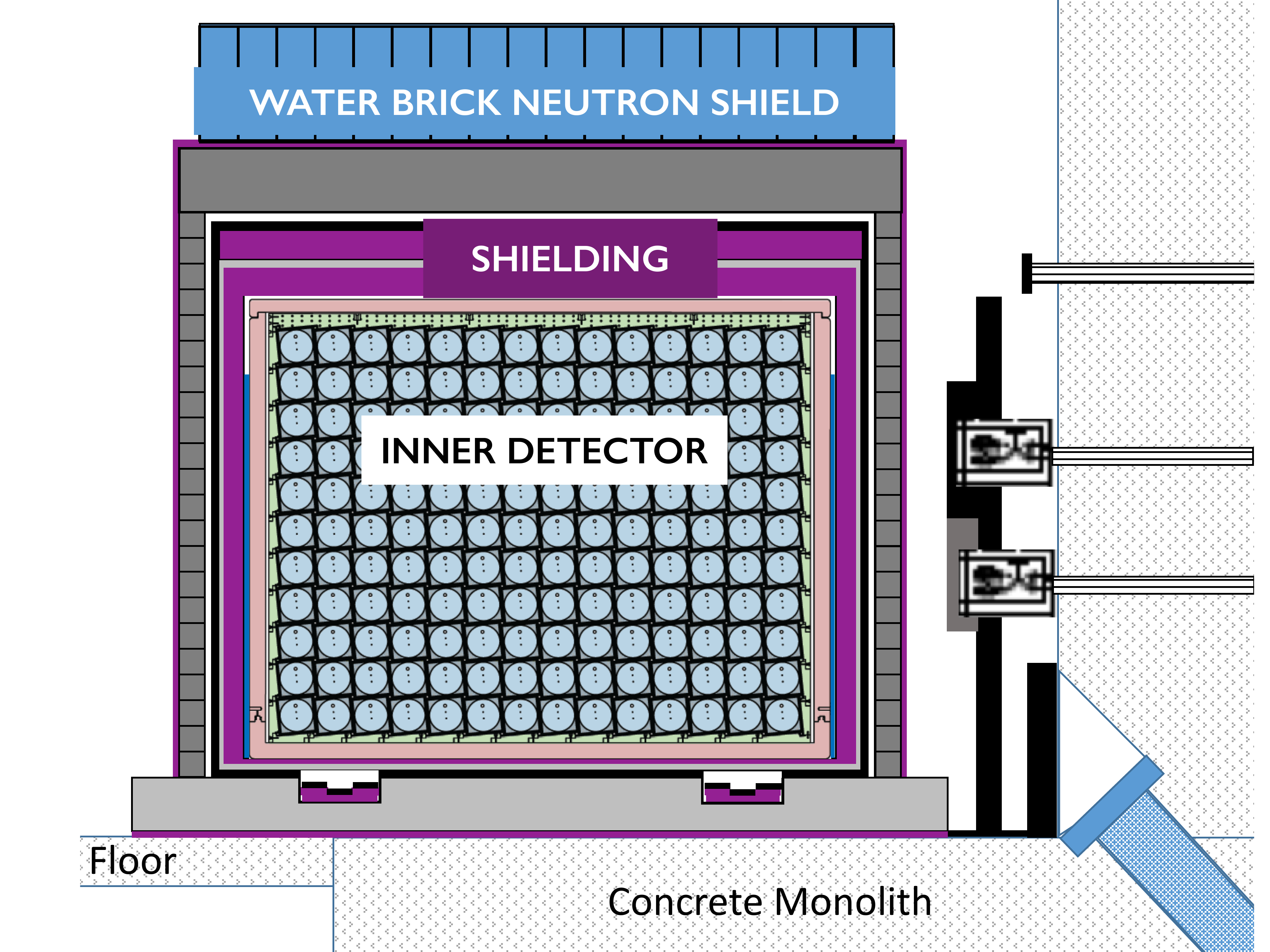}
\caption{Cross-sectional view of the PROSPECT AD at the HFIR site. Also shown in the figure is the shielding package including lead, polyethylene, and water bricks. Additionally, the lead wall adjoining the reactor pool wall is also shown on the right.}
\label{fig:Detector}
\end{figure}
PROSPECT is an experiment designed with the primary goals to search for oscillation suggested by the flux anomaly and precisely measure \uFive~spectrum to address the spectrum anomaly~\cite{PROSPECTNIM}.
The source of \nuebar~for PROSPECT is the High Flux Isotope Reactor, an HEU research reactor located at Oak Ridge National Laboratory.
The combination of HEU fuel and the short reactor cycles constrains the contribution of reactor \nuebar~from \uFive~to greater than 99\% throughout the reactor cycle. 

The PROSPECT Antineutrino Detector~(AD) is an optically segmented detector with $^6$Li-loaded liquid scintillator~(LiLS) as the target for Inverse Beta Decay~(IBD) interaction~\cite{PROSPECTNIM}.
The segmentation of the detector is achieved by the use of custom-designed optical separators~\cite{PROSPECT_OG} which guide the scintillation light to be collected by the PMTs placed at either ends of a segment. 
Consisting of a total of 11x14 segments~(154) with each individual segment 119 cm long and a cross-section of 14.6 cm x 14.6 cm, segmentation provides a capability to deploy sources within the detector.

PROSPECT AD is located at a closest distance of $\sim$7m to the reactor enabling a high statistics measurement and the ability to search for high-frequency oscillations, but it also poses a challenge of high reactor-related and cosmogenic backgrounds.  
The PROSPECT collaboration has done extensive background characterization~\cite{Prospect_BG} of the reactor site and has developed several active and passive background rejection techniques to mitigate backgrounds.
The primary source of IBD-mimicking backgrounds -- cosmogenic fast neutrons interacting with detector and surrounding materials -- are reduced by the use of an optimized top-heavy shielding package consisting of water bricks, borated polyethylene, and lead.
The high rates of primarily $\gamma$-backgrounds from reactors are reduced passively by the use of a purpose-built lead wall.
The segmentation of the detector provides 3D-event reconstruction capability which in turn provides a way to apply topological and fiducialization cuts.
The liquid scintillator used in PROSPECT, EJ-309 is PSD-capable and provides the ability to separate electron-recoils from nuclear-recoils which is leveraged to reject correlated multi-neutron events as well as uncorrelated gamma coincidences. 

\begin{figure}[!htb]
\centering
\includegraphics[width=0.49\textwidth]{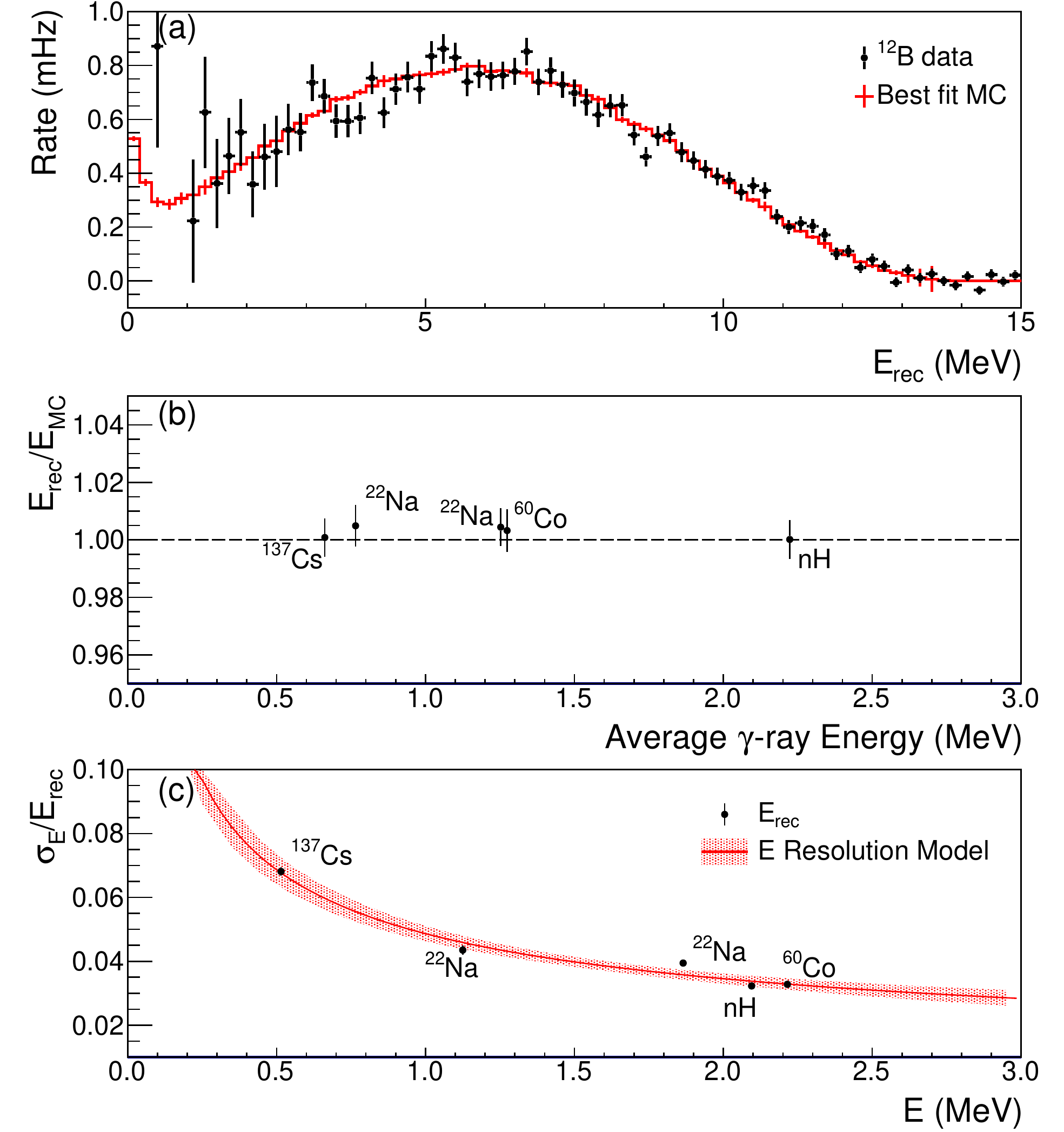}
\includegraphics[width=0.49\textwidth]{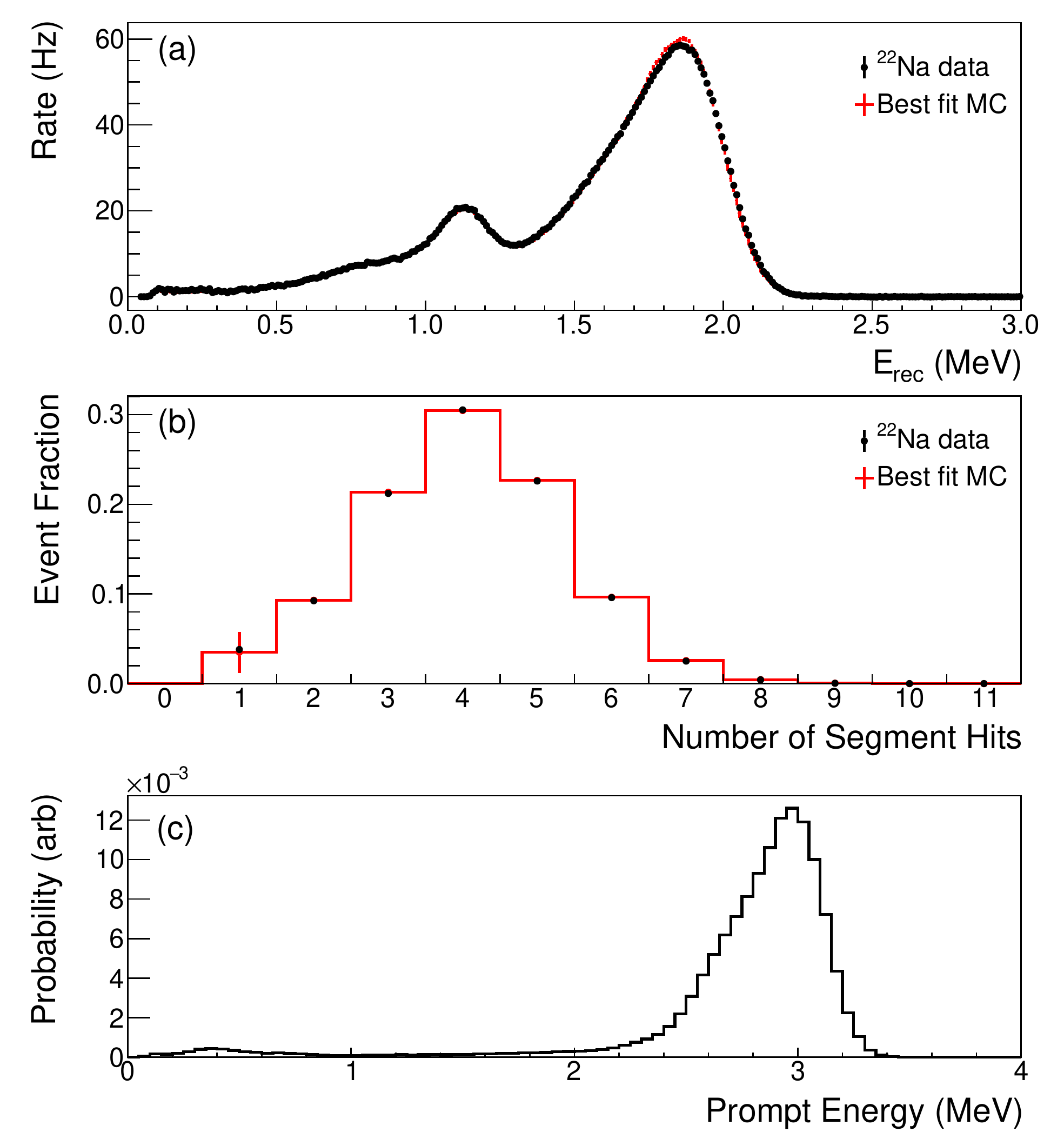}
\caption{\textit{Left} (a): The $^{12}$B energy spectrum compared with the corresponding best-fit PG4 spectrum. (b): Ratio of the reconstructed to PG4 average energies of $\gamma$ calibration sources. (c): Energy resolution as a function of reconstructed energy for various calibration sources.
\textit{Right} (a): Comparison between reconstructed and best-fit PG4 energy spectrum of the $^{22}$Na spectrum. (b): Segment multiplicity of $^{22}$Na events. (c): Reconstructed energy spectrum of 4 MeV \nuebar. The distribution is asymmetric because of escaping annihilation $\gamma$s.}
\label{fig:Response}
\end{figure}
The calibration and energy reconstruction of the PROSPECT detector is done in multiple steps.
The integral of waveforms collected by the PMTs are converted to position-corrected energy. 
The relative energy scale between segments are constrained by $^6$Li capture peak which has small spatial and temporal energy deposition while the absolute energy scale is defined based on the fits performed on the calibration data to Monte-Carlo simulations.
The calibration of the detector is done by using the deployed $\gamma$-calibration sources $^{22}$Na, $^{60}$Co, and $^{137}$Cs; ambient 2.2 MeV $\gamma$s from neutron capture on hydrogen and $\beta$~spectrum from cosmogenic neutron-capture on $^{12}$C by $^{12}$C(n,p)$^{12}$B and subsequent beta decay of $^{12}$B.
The full-detector energy spectra from these sources and the segment multiplicity of the $^{22}$Na are simultaneously fit to PROSPECT Geant4-based Monte Carlo simulations~(PG4) of the respective source spectra and multiplicity distributions. 
The fits are performed by varying absolute energy scale, Birks parameter, Cherenkov parameter and energy resolution. 
Figure \ref{fig:Response} shows the agreement between the calibration data and PG4 simulations.

\section{Spectrum Measurement}
\begin{figure}[!htb]
\centering
\includegraphics[width=\textwidth]{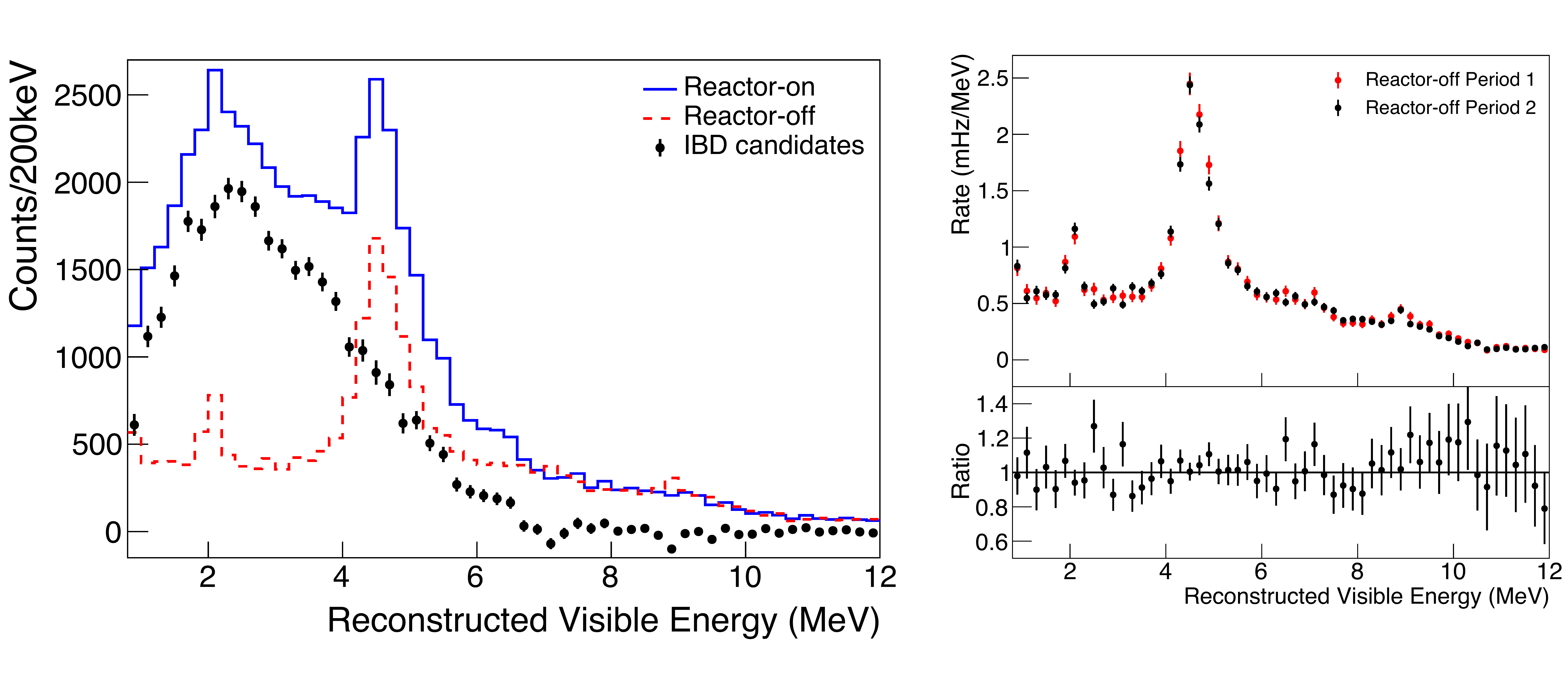}
\caption{\textit{Left}: Correlated reactor on~(blue) and reactor off~(red) spectra after accidental background subtraction and reconstructed prompt energy spectrum with statistical errors bars.
\textit{Right}: Scaled time-separated cosmogenic background spectra demonstrating the stability of the PROSPECT detector over time.}
\label{fig:spectrum}
\end{figure}

For this analysis two interleaved datasets of reactor on~(off) periods are used summing to a total of 40.3~(37.8) days of exposure respectively~\cite{bib_PROSPECT_Spectrum}.
After applying analysis cuts, this dataset corresponds to a total of 70811 $\pm$ 267~(20036 $\pm$ 145)~correlated reactor on~(off) events, 20534 $\pm$ 16~(1436 $\pm$ 4) reactor on~(off) accidental events resulting in 31678 $\pm$ 304 \nuebar~events after atmospheric pressure-corrected background subtraction with a signal-to-background of 1.7:1.
Fig.~\ref{fig:spectrum} shows the correlated reactor on, off, and IBD spectrum as well as reactor off correlated rates for two distinct data periods.

\begin{figure}[!htb]
\centering
\includegraphics[width=0.6\textwidth]{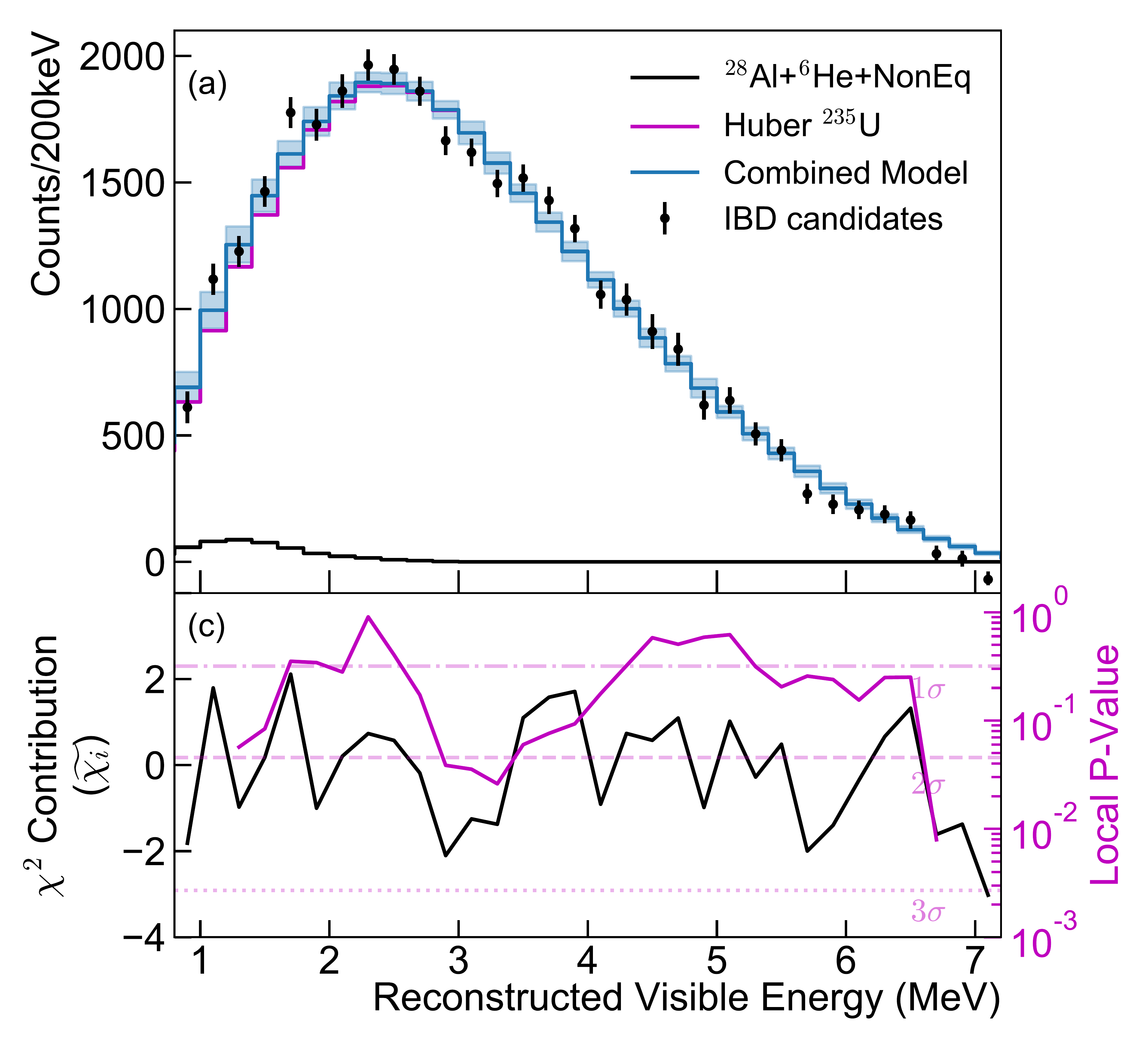}
\caption{PROSPECT's Reconstructed prompt energy spectrum in comparison with predicted prompt energy spectrum are shown in the top panel. The combined model~(blue) is estimated by summing the contributions from Huber model~(pink) non-fission as well as non-equillibrium isotopes~(black). Also shown in the bottom panel are the bin-by-bin signed $\tilde{\chi}$ contributions and p-values from the sliding window test.}
\label{fig:CombinedComparison}
\end{figure}
A quantitative comparison between the measured spectrum and the Huber model is done by the use of a $\chi^2$ test-statistic defined by 
\begin{equation}
    \chi^2_{min}=\mathbf{\Delta}^{T}\textbf{V}^{-1}\mathbf{\Delta},
    \label{chi2}
\end{equation}
where $\mathbf{\Delta}_{i}$ is the difference between measured and predicted neutrino counts in bin $i$ and \textbf{V} incorporates signal and background statistical and systematic uncertainties and their bin-to-bin correlations.
To construct the predicted rates the  $^{235}U$ Huber neutrino spectrum is added to the non-fission isotope contributions from the beta decays of $^{28}$Al and $^{6}$He and contributions from the non-eq isotopes~\cite{conant_thesis}. 
This summed neutrino energy spectrum is then passed through the PG4-generated detector response model to obtain the predicted prompt energy spectrum. 
The $\chi^2$ is minimized over a free-floating nuisance parameter is incorporated into the predicted neutrino counts corresponding to the overall measured neutrino count effectively yielding a shape-only fit.

The predicted and measured reconstructed energy spectra are shown in the Fig.~\ref{fig:CombinedComparison}. 
The shape-only fit yields a $\chi_{min}^2$/NDF of 51.4/31 and demonstrates an overall bad fit with a one-side p-value of 0.01.
The bottom panel of the figure shows a signed test-statistic defined by 
\begin{equation}
    \tilde{\chi_i}=\frac{N^{obs}_{i}-N^{pred}_{i}}{|N^{obs}_{i}-N^{pred}|}\sqrt{\chi^2_{original}-\chi^2_{i,new}}
    \label{chi2tilde}
\end{equation}
where the $\chi^2_{original}$ is the value calculated from Eq.~\ref{chi2} and $\chi^2_{i,new}$ is calculated by incorporating an additional free-floating nuisance parameter corresponding to the content in the $i^{\textrm{th}}$ bin.
This metric which shows the bin-to-bin significance of deviation displays no notable deviations except for the final bin.
Another procedure called sliding window test following the methodology employed in Ref.~\cite{An:2016srz} is performed by letting consecutive bins over a region of 1 MeV be free-floating.
The results displayed in~Fig.~\ref{fig:CombinedComparison} shows two regions over the energy range that disagree with the Huber model at more than 2$\sigma$.     

\begin{figure}[!htb]
\centering
\includegraphics[width=0.8\textwidth]{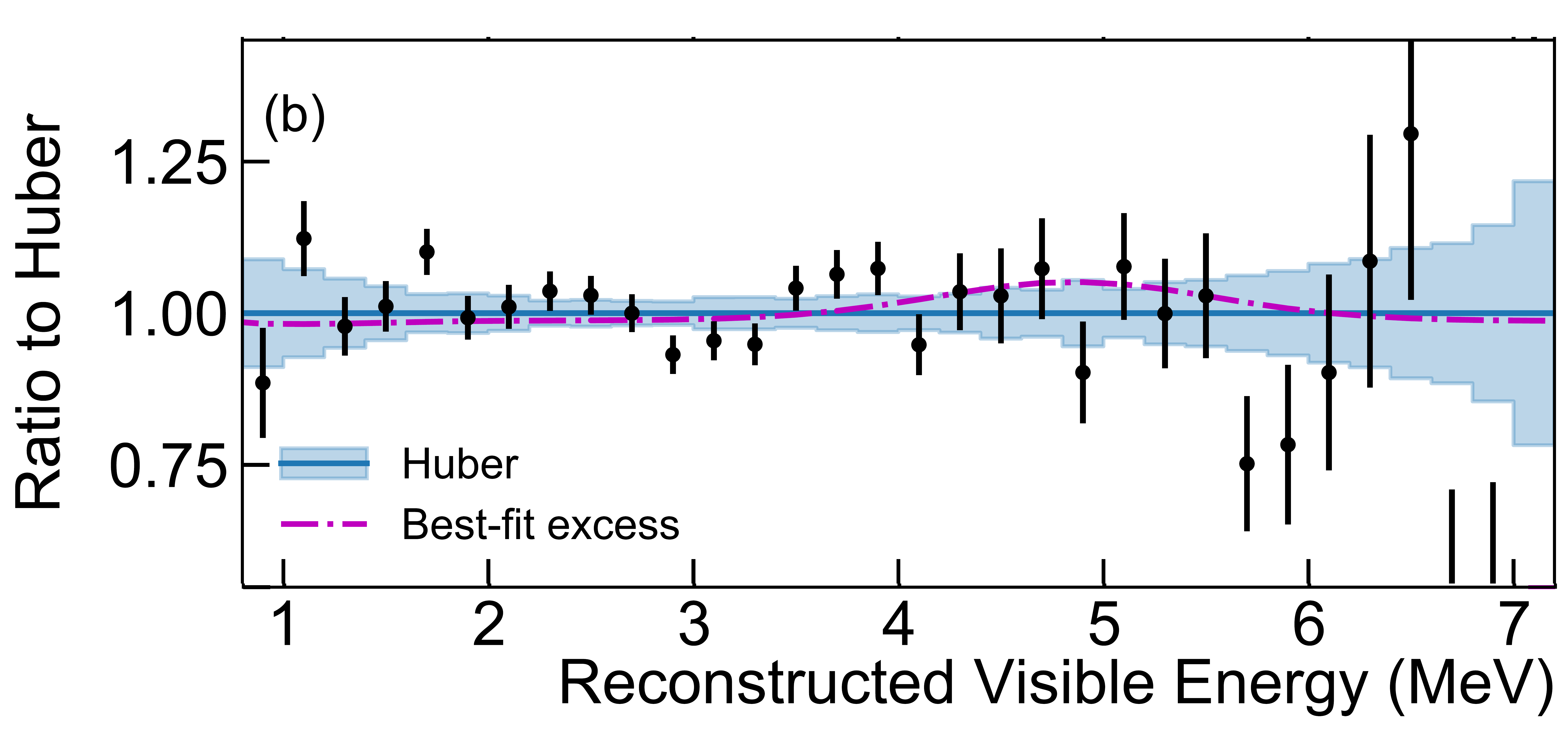}
\caption{The ratio of measured spectrum to the predicted spectrum. Black error bars show statistical errors while the blue bands show systematic uncertainties. The dotted line in pink shows the best-fit ad-hoc model}
\label{fig:Adhoc_spect}
\end{figure}
An ad-hoc model is constructed by fitting the difference between Daya Bay measured spectrum and Huber-Mueller models to a Gaussian to test if the LEU fluctuations in 5-7 MeV region originate from \uFive. 
The mean and sigma from the fit is then added to PROSPECT data and a $\chi^2$ as defined in Eq.~\ref{chi2} is calculated for varying values of normalization $n$. 
A fit with a normalization $n=1$ corresponds to the Daya Bay size local deviation while $n=0$ corresponds to the Huber model.
PROSPECT data agrees with both the models with the best-fit normalization found at 0.69 $\pm$ 0.53 as shown in Fig.~\ref{fig:Adhoc_spect}.
A fit with a normalization of $n=1.78$ corresponding to \uFive~being solely responsible for the Daya Bay local deviation is disfavoured at 2.1$\sigma$.

\section{Conclusions}
PROSPECT built an on-surface detector close to a reactor and performed the first modern measurement of high-statistics antineutrino spectrum from an HEU reactor with a S:B of 1.7:1.
A shape-only comparison between PROSPECT data and Huber model has a $\chi^2$/NDF exhibiting a poor overall fit. 
With the current statistics-limited data an ad-hoc model corresponding to the local deviation at 5-7 MeV arising purely from \uFive~is disfavored at 2.1$\sigma$.

\section*{Acknowledgements}
This material is based upon work supported by the U.S. Department of Energy Office of Science and the Heising-Simons Foundation. Additional support is provided by Illinois Institute of Technology, LLNL, NIST, ORNL, Temple University, and Yale University. We gratefully acknowledge the support and hospitality of the High Flux Isotope Reactor, managed by UT-Battelle for the U.S. Department of Energy.

\bibliography{citations.bib}
 
\end{document}